\documentclass[twocolumn]{aastex61}
\pdfoutput=1 
\usepackage{amsmath,amstext}
\usepackage[T1]{fontenc}
\usepackage{apjfonts} 
\usepackage[figure,figure*]{hypcap}
\usepackage[utf8]{inputenc}

\def\mgb{\rm Mg\,\textit b}
\def\hb{H$\beta$}
\def\fei{\rm Fe\,5270}
\def\feii{\rm Fe\,5335}
\def\fe{$\langle \rm Fe \rangle$}
\def\afe{[$\alpha$/Fe]} 
\def\mgfe{\rm[MgFe]$^\prime$}

\shorttitle{AASTeX 6.1 Template}
\shortauthors{Kim et al.}

\begin{document}
\title{Compact elliptical galaxies in different local environments: a mixture of galaxies with different origins?}

\correspondingauthor{Suk Kim \& Soo-Chang Rey}
\email{star4citizen@gmail.com, screy@cnu.ac.kr}

\author{Suk Kim}
\affiliation{Department of Astronomy and Space Science, Chungnam National University, Daejeon 34134, Republic of Korea; star4citizen@gmail.com, screy@cnu.ac.kr}
\affiliation{Research Institute of Natural Sciences, Chungnam National University, Daejeon 34134, Republic of Korea}
\affiliation{Center for Galaxy Evolution Research, Yonsei University, Seoul 03722, Republic of Korea}

\author{Hyunjin Jeong}
\affiliation{Korea Astronomy $\&$ Space Science Institute, Daejeon 34055, Republic of Korea}

\author{Soo-Chang Rey}
\affiliation{Department of Astronomy and Space Science, Chungnam National University, Daejeon 34134, Republic of Korea; star4citizen@gmail.com, screy@cnu.ac.kr}

\author{Youngdae Lee}
\affiliation{Department of Astronomy and Space Science, Chungnam National University, Daejeon 34134, Republic of Korea; star4citizen@gmail.com, screy@cnu.ac.kr}
\affiliation{Research Institute of Natural Sciences, Chungnam National University, Daejeon 34134, Republic of Korea}

\author{Jaehyun Lee}
\affiliation{Korea Institute for Advanced Study, Seoul 02455, Republic of Korea}

\author{Seok-Joo Joo}
\affiliation{Department of Astronomy and Space Science, Chungnam National University, Daejeon 34134, Republic of Korea; star4citizen@gmail.com, screy@cnu.ac.kr}
\affiliation{Research Institute of Natural Sciences, Chungnam National University, Daejeon 34134, Republic of Korea}

\author{Hak-Sub Kim}
\affiliation{Korea Astronomy $\&$ Space Science Institute, Daejeon 34055, Republic of Korea}

\begin{abstract}
We present the stellar populations of 138 compact elliptical galaxies (cEs) in the redshift range of $z < 0.05$ using the Sloan Digital Sky Survey (SDSS) DR12. Our cEs are divided into those with [cE(w)] and without [cE(w/o)] a bright ($M_{r} < -21$ mag) host galaxy. We investigated the stellar population properties of cEs based on the Lick line indices extracted from SDSS spectra. cE(w)s show [Z/H] and \afe\ distributions skewed toward higher values compared to those of the cE(w/o)s. No statistically significant difference in age distribution was found between the cE(w)s and cE(w/o)s. In the mass-metallicity distribution, cE(w)s deviate from the relation observed for early-type galaxies at a given stellar mass, whereas cE(w/o)s conform to the relation. Based on the different features in the stellar populations of cE(w)s and cE(w/o)s, we can propose two different cE formation channels tracing different original masses of the progenitors. cE(w)s would be the remnant cores of the massive progenitor galaxies whose outer parts are tidally stripped by a massive neighboring galaxy (i.e., nurture origin). In contrast, cE(w/o)s are likely the faint end of early-type galaxies maintaining in-situ evolution in an isolated environment with no massive galaxy nearby (i.e., nature origin). Our results reinforce the propositions that cEs comprise a mixture of galaxies with two types of origins depending on their local environment.

\end{abstract}


\section{Introduction}

Compact elliptical galaxies (cEs) are relatively rare objects in the local universe, characterized by very small effective radii (a few hundred pc), low stellar masses ($10^{8}$ -- $10^{10}M_{\odot}$), and high central surface brightnesses \citep{Faber73,Norris2014}.
Therefore, cEs are located in the low-mass end of massive early-type galaxies and differ from the diffuse low-mass galaxies such as dwarf early-type galaxies (dEs) in their mass-size distribution. While several scenarios have been suggested \citep{Faber73,Bekki2001,Martinovic2017,Du2019}, questions regarding the origin of cEs remain unanswered.

Most cEs deviate from the mass-metallicity relation observed in classical early-type galaxies in that cEs are more metal-rich than galaxies of comparable masses; however, they have metallicities appropriate to more massive galaxies \citep{chilingarian2015,Janz2016}. Furthermore, cEs are most likely to be associated with an adjacent massive host galaxy in galaxy clusters or groups \citep{Norris2014,chilingarian2015,Janz2016}. These facts suggest a formation scenario in which cEs are the tidally stripped remnants of larger, more massive galaxies \citep{Faber73,Bekki2001,Choi2002,Graham2002}. According to this scenario, an early-type disk galaxy with a compact bulge loses a large fraction of its initial disk mass through dissipative tidal interactions with a more massive host galaxy and only the central bulge component survives \citep[e.g.,][]{Bekki2001,Chilingarian2009}. The outcome is a metal-rich and low-mass cE that is in line with observational results \citep{Norris2014,chilingarian2015,Janz2016}. Indeed, the discovery of tidal streams around a few cEs can be considered as a direct evidence for cE formation through tidal stripping \citep{Huxor2011,Paudel2013,chilingarian2015,Ferre2018}. 
Further, supporting evidence of the cE formation through stripping is that cEs should contain a central black hole (BH) with mass appropriate for becoming a massive progenitor \bibpunct[ ]{(}{)}{;}{a}{}{;}\citep[see][for details of the observational hints of central BHs existing in cEs]{Kormendy1997,VanDerMarel1997,Forbes2014,Paudel2016}.

An alternative cE formation scenario proposes that cEs are instead the natural extension of classical luminous elliptical galaxies to lower luminosities. This is supported by the fact that cEs follow the scaling relations of giant elliptical galaxies at the low-mass end \bibpunct[; ]{(}{)}{;}{a}{}{;}\citep[e.g.,][see also \citealp{Saulder2015} for more massive cEs with stellar masses in the range of 10$^{10}$ - $10^{11}M_{\odot}$]{Wirth1984,Kormendy2009,Kormendy2012}.
\citet{Kormendy2012} argue against the stripping scenario since not all cEs are companion galaxies of bright galaxies. The discovery of isolated cEs is conclusive evidence that cEs may not be formed by appreciable environmental effects, suggesting an alternative channel to the tidal stripping for the cE formation \citep{Huxor2013,Paudel2014}. It has been suggested that isolated cEs might have originated through an earlier merger between smaller objects, as in typical massive elliptical galaxies \citep[e.g.,][]{Kormendy2009,Paudel2014}.
 
From their large sample of cEs identified in various environments including 11 isolated cEs, \citet{chilingarian2015} found that dynamical and stellar population properties of isolated cEs might be similar to those in more dense environments. They suggested that isolated cEs are originally formed by tidal stripping in clusters or groups and then ejected from the host galaxy via three-body encounters. Therefore, this finding has simplified the formation scenarios of cEs, eliminating the need for an additional formation channel for the isolated cEs.

In the context of current galaxy formation scenarios, most compact galaxies formed at high-z will evolve into classical massive galaxies through hierarchical merging, rather than retaining their compact morphology by $z=0$ \citep{Damjanov2009,DeLaRosa2016,Huang2016,Rodriguez2016}. However, according to cosmological simulations, some compact galaxies formed at high-z are likely to survive in the local universe \citep[e.g.,][]{Wellons2016,Martinovic2017}. In this case, these galaxies might not have acquired {\it ex situ\/} stellar mass over time owing to the lack of mergers with other galaxies. Such intrinsic cEs are found as low-mass ellipticals formed at very early cosmic epochs; they are expected to follow the same trend as that of the compact cores of giant ellipticals for the mass-metallicity relation in terms of their predicted population properties.

cEs are ubiquitous in a wide variety of environments \citep{Chilingarian2009,chilingarian2015,Norris2014,Janz2016,Zhang2017,Ferre2018}. While most cEs exist in a dense cluster environment, some are also found in loose environments of galaxy groups and fields. All previous studies have concentrated on large-scale, global environments containing cEs \citep[e.g.,][]{Norris2014,chilingarian2015,Janz2016}. However, small-scale environments of galaxies also have an impact on the various properties of the galaxies \citep[e.g.,][]{Park2008,Park2009,Geha2012,Robotham2014,Alpaslan2015}. Therefore, it is important to study the dependence of cE properties on the local environment relative to their neighboring massive galaxy.

In this regard, most recently, \citet{Ferre2018} studied various properties of 25 cEs at different evolutionary stages according to the possible interactions with their massive neighbor galaxy. Although the number of cEs has grown over the last few years, a more extensive and homogeneous sample of cEs in different local environments is required \citep[e.g.,][]{Chilingarian2009,chilingarian2015}. This would allow a more systematic investigation of properties of cEs to obtain further insights regarding cE formation. Hence, in this study, we searched for a large sample of cEs at $z <$ 0.05 from the Sloan Digital Sky Survey (SDSS) DR12. We studied the stellar population properties of cEs depending on the local environment related to their host galaxy. In particular, we focused on comparing stellar population properties between two cE subsamples, namely those associated or not associated with a close host galaxy. The implications for different properties of cE subsamples are discussed in terms of rivaling cE formation scenarios.

This paper is organized as follows. In Section 2, we describe the cE sample construction. In Section 3.1, we present the environmental parameterization of cEs and the classification of cEs into subsamples with or without nearby massive host galaxy. In Section 3.2, we compare the stellar population properties of the two subsamples. In Section 3.3, we present the mass-metallicity distributions of the cEs. Finally, we discuss the formation scenarios of cEs and summarize the results in Section 4. Throughout this study, we assumed the cosmological parameters to be $\Omega_{m} = 0.3$, $\Omega_{\Lambda} = 0.7$, and $h_{0} = 0.73$.

\section{SAMPLE SELECTION}
\begin{figure}
\vspace{3mm}
  \includegraphics[width=3.3in]{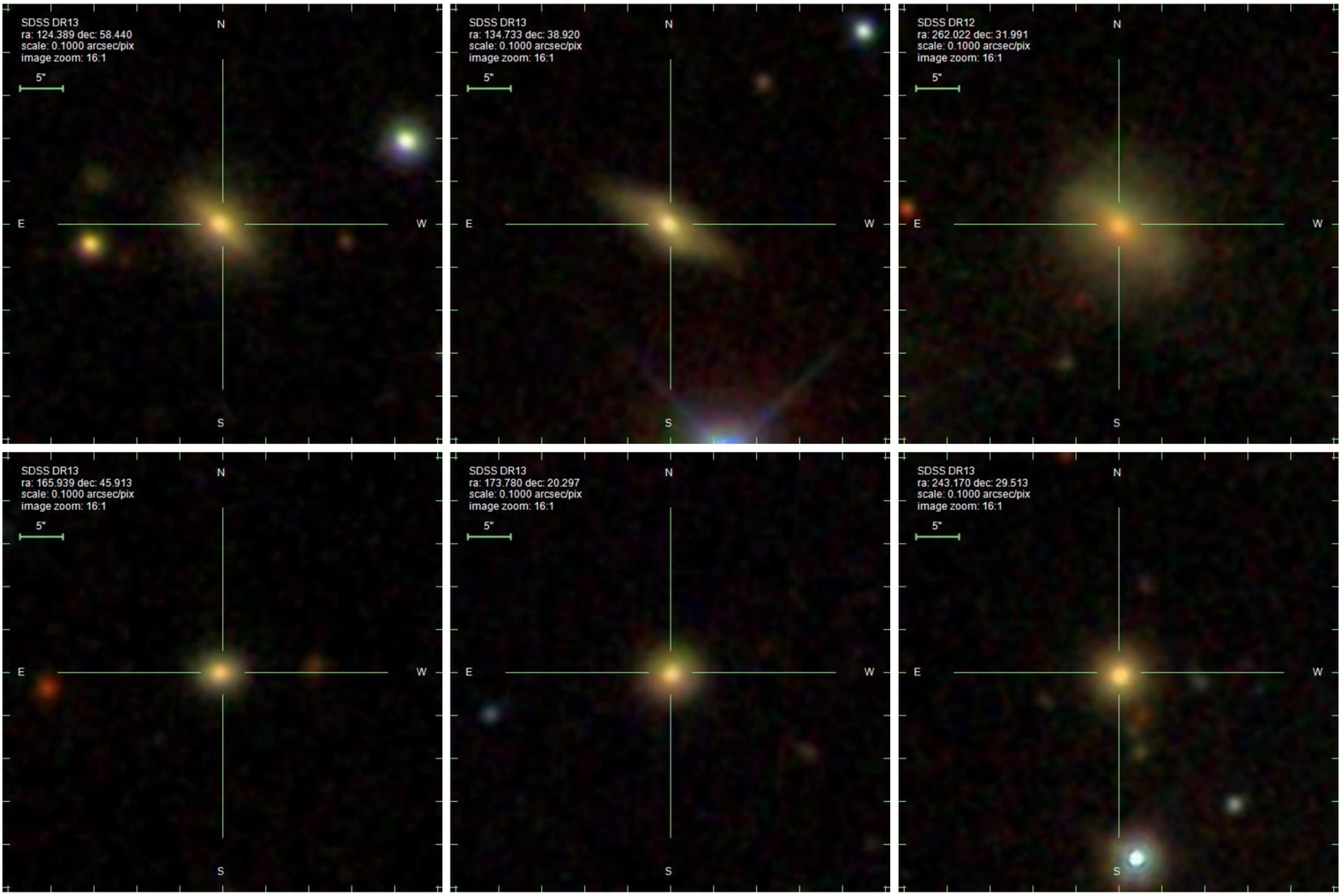}
  \caption{Examples of discarded galaxies with underlying disks or irregular shapes (top panels) from initial cE candidates and selected as final cE candidates (bottom panels) by visual inspection.}
  \label{fig:Eximg}
\end{figure}

While cEs show luminosities similar to bright dEs, the effective radii of cEs are smaller than those of dEs. Moreover, cEs also have larger velocity dispersions compared to dEs at the same luminosity \citep{Kormendy1985}. Following the selection criteria of \citet{chilingarian2015}, we created a list of cE candidates in the redshift range of $z <$ 0.05 using the SDSS DR12. The distances of these galaxies are estimated using radial velocities extracted from the SDSS based on the linear relationship between the radial velocity and distance. We initially chose low-luminosity ($M_{g} > -18.7$ mag) galaxies with a small Petrosian effective radius ($R_{\rm eff, petro}$ $<$ 600 pc) or those remaining spatially unresolved in SDSS images. We considered objects redward of the $-3\sigma$ deviation from the universal red sequence in the $g-r$ vs. $M_{r}$ color-magnitude relation constructed using all galaxies of the SDSS DR12. In particular, we selected galaxies with high-velocity dispersion ($\sigma$ $>$ 60 km s$^{-1}$) to accurately exclude dEs of similar brightnesses. The selected cE candidates included a total of 2,352 galaxies. Three members of our team (S.K., H.J., $\&$ Y.L.) independently performed visual inspection of the images of all galaxies. The morphology of each galaxy was finalized if the classification of two or more classifiers agreed. Most galaxies show an underlying disk or central irregularity (see top panels of Figure~\ref{fig:Eximg}). We secured 403 cE candidates with only compact elliptical shapes (see bottom panels of Fig.~\ref{fig:Eximg}). We further excluded 21 candidates with possible ongoing star-formation activities that occupy the region of star-forming galaxies in the Baldwin-Phillips-Terlevich diagram \citep{Baldwin1981}.

We used the $R_{\rm eff, petro}$ provided by the SDSS DR12 for selection of the cE candidates. However, the $R_{\rm eff, petro}$ does not consider the seeing effect. The effective radius would be buried in the seeing size when the actual effective radius is smaller than the seeing size of the SDSS image. Hence, we redetermined the sizes of galaxies from SDSS $r$-band images using GALFIT \citep{Peng02}, a two-dimensional galaxy-fitting code. For each galaxy image, GALFIT creates a convolution of a S\'{e}rsic model with the point spread function from a set of point sources within the image field. GALFIT finally provides the model effective radius ($R_{\rm eff, model}$) of all cE candidates from a single-component S\'{e}rsic fit determined by comparing the convolved image with the SDSS image of the galaxy \citep[see also][]{Trujillo2006,Huxor2013}.

\begin{figure}
\vspace{3mm}
  \includegraphics[width=3.3in]{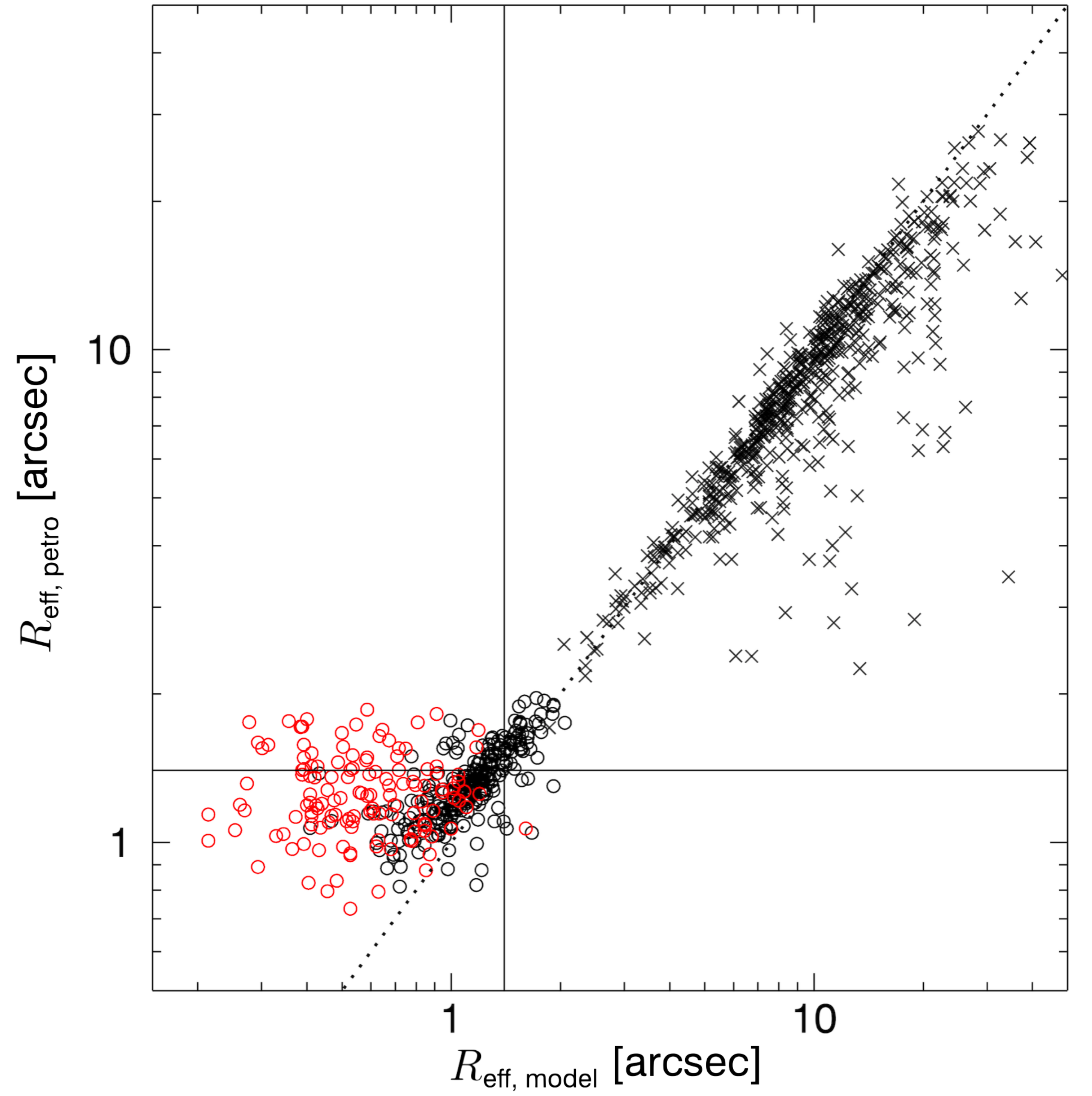}
  \caption{Comparison between SDSS Petrosian effective radius ($R_{\rm eff, petro}$) and S\'{e}rsic model effective radius ($R_{\rm eff, model}$) returned from GALFIT of cE candidates (circles) and early-type galaxies in the Virgo cluster (crosses). The dotted line is the 1:1 relation to guide the eye. The vertical and horizontal solid lines show the typical seeing size of the SDSS $r$-band image ($\sim$1.4 arcsec). Red circles denote 138 cEs with sizes less than 600 pc as the final sample for our analysis.}
  \label{fig:Re}
\end{figure}

Figure~\ref{fig:Re} shows a comparison between $R_{\rm eff, petro}$ and $R_{\rm eff, model}$ of the galaxies. The crosses are early-type galaxies in the Virgo cluster selected from the Extended Virgo Cluster Catalog \citep{Kim2014}. Their $R_{\rm eff, petro}$ values were also adopted from the SDSS DR12, and the $R_{\rm eff, model}$ was derived using GALFIT. Early-type galaxies in the Virgo cluster mostly follow the expected correlation between $R_{\rm eff, petro}$ and $R_{\rm eff, model}$ values. However, a significant fraction of our sample (circles) deviate from the 1:1 relation (dotted line). When $R_{\rm eff, model}$ was smaller than the seeing size of the SDSS r-band image ($\sim$1.4 arcsec \footnote{https://classic.sdss.org/dr7/}, see solid lines in Fig~\ref{fig:Re}), $R_{\rm eff, petro}$ values are larger than $R_{\rm eff, model}$ values. Therefore, we used the $R_{\rm eff, model}$ values of all cE candidates to determine their sizes. 
Finally, we selected 138 cEs with sizes less than 600 pc as the final sample for our analysis (see red circles in Fig. 2).

\begin{figure}
\vspace{3mm}
  \includegraphics[width=3.4in]{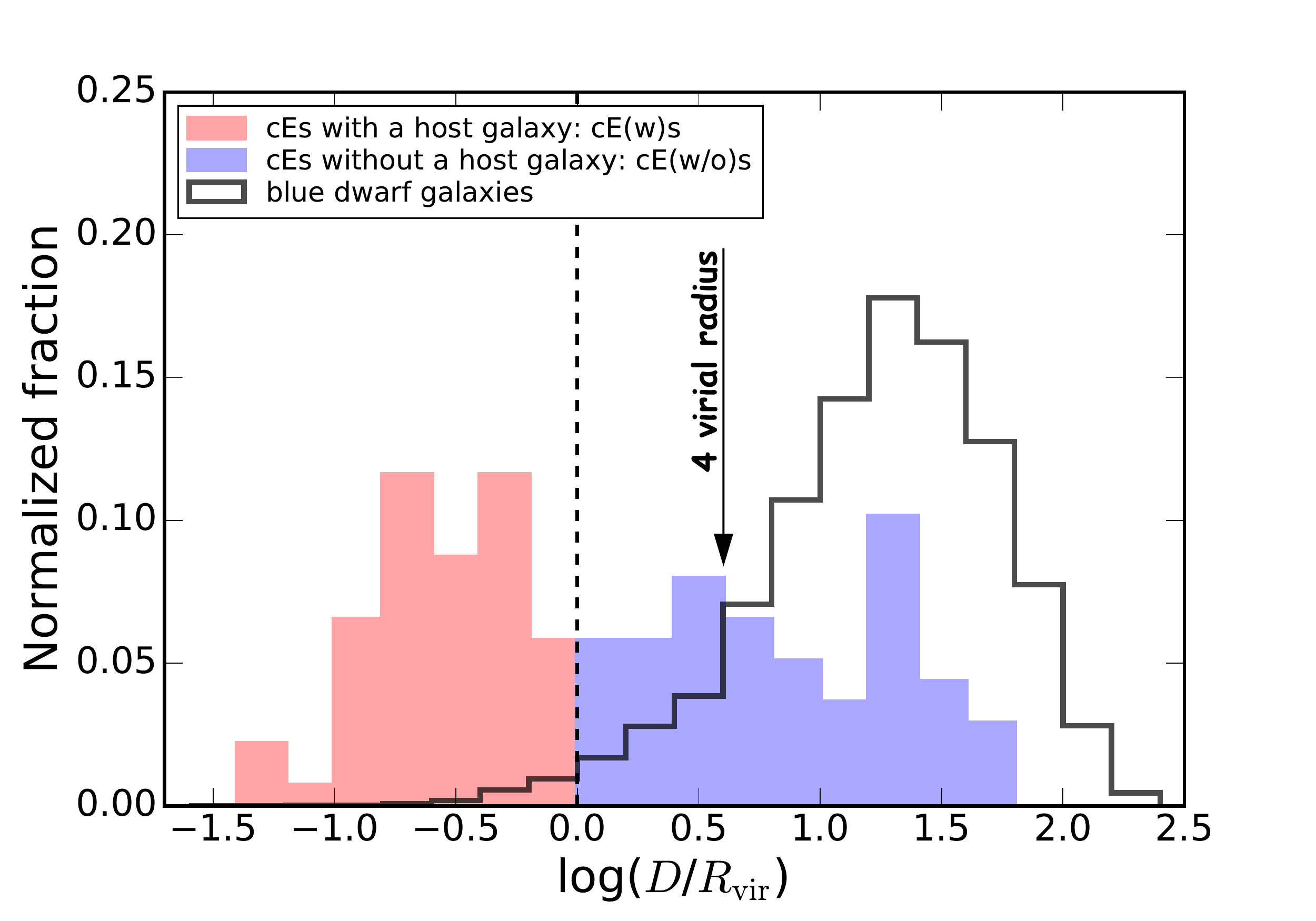}
  \caption{Distribution of the projected distance of cEs from their nearest luminous galaxy in the unit of its virial radius $R_{\rm vir}$. Red and blue histograms are for cE(w)s and cE(w/o)s, respectively which are divided by one $R_{\rm vir}$ of the nearest luminous galaxy (dotted line). The black open histogram denotes the distribution of possible blue, star-forming dwarf galaxies in the same luminosity range as that of cE sample. It is worth noting that most cE(w/o)s are located in lower-density local environments, similar to blue dwarf galaxies.}
  \label{fig:Envir}
\end{figure}

\begin{figure}
\vspace{0mm}
  \includegraphics[width=3.7in]{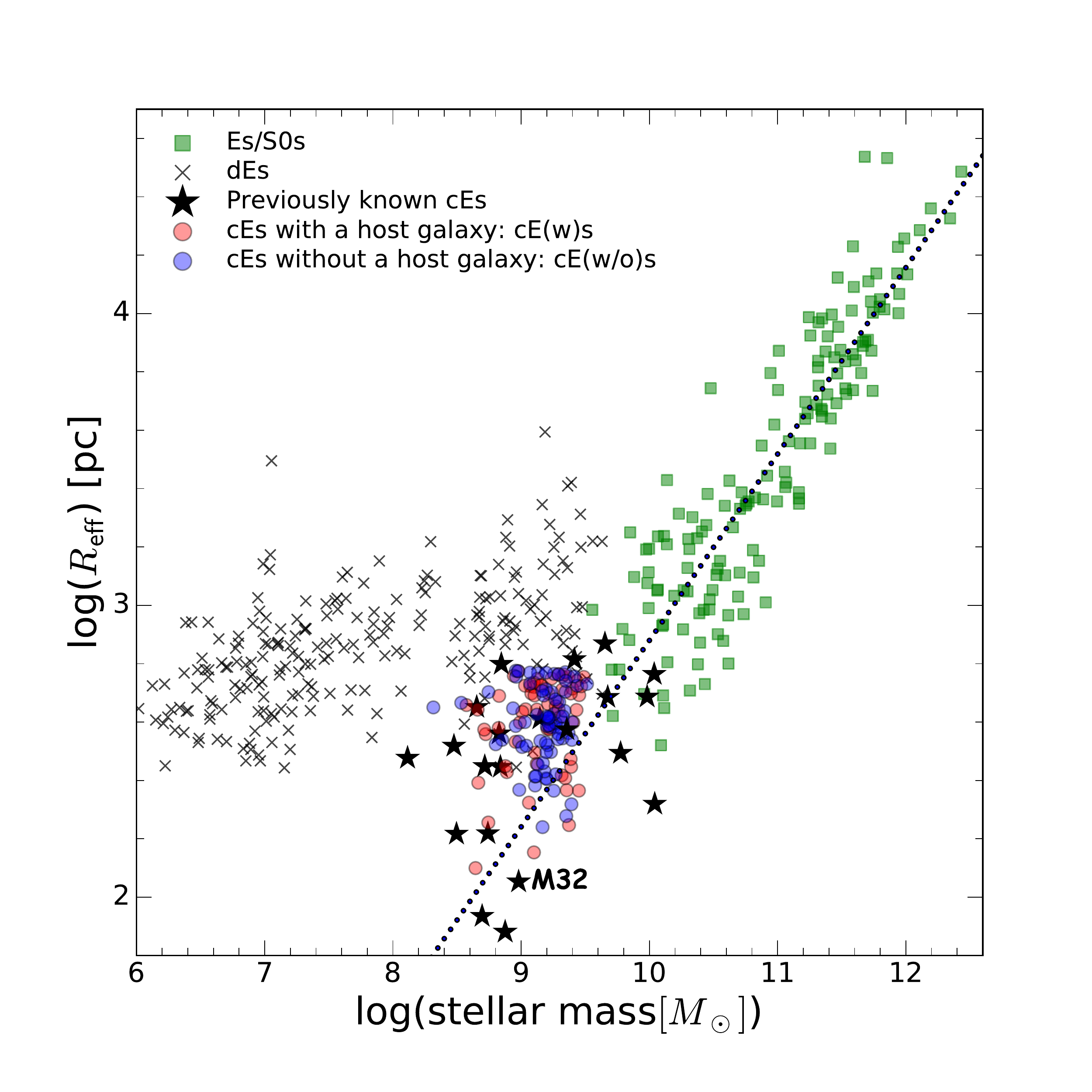}
  \caption{Effective radius ($R_{\rm eff}$) vs. stellar mass of various types of galaxies. Circles represent our cE sample divided into different local environments; red and blue circles indicate cE(w)s and cE(w/o)s, respectively. Green boxes and crosses denote early-type galaxies (E/S0s) and dEs, respectively, extracted from literature \citep{Bender1993,Ferrarese2006,Misgeld2008,Misgeld2009}. Stars denote previously known cEs \citep{Bender1993,Norris2014,Guerou2015}. M32 - a "classical" cE as a very close satellite of M31 - is also labeled. The dotted line represents the linear best fit for E/S0s.}
    \label{fig:MS}
\end{figure}

\section{Results}

\subsection{Environments of Compact Ellipticals}

We quantified the small-scale, local environment of our cE sample using the projected distances from each cE to the nearest luminous (M$_{r}$ $< -$21.0 mag) galaxies that have velocity differences less than 500 km s$^{-1}$ with respect to the cE. We then defined the nearest luminous galaxy and calculated its virial radius ($R_{\rm vir}$) using the relation between the effective radius (R$_{\rm eff}$) and $R_{\rm vir}$, given as $R_{\rm eff} = 0.015R_{\rm vir}$ \citep{Kravtsov2013}, where the $R_{\rm eff}$ of the nearest luminous galaxy was taken from the SDSS DR12. We divided our cEs into those inside and outside of one $R_{\rm vir}$ of the nearest luminous galaxy, which were then classified as those with [cE(w)s] and without [cE(w/o)s] a host galaxy, respectively. Our cE sample contains 65 cE(w)s and 73 cE(w/o)s.
This division is based on the assumption that the cE could be physically associated with the host galaxy when the cE is located within one $R_{\rm vir}$ of the host galaxy.

Figure~\ref{fig:Envir} presents the distribution of the projected distance of cEs relative to their nearest luminous galaxy in units of its $R_{\rm vir}$. The red and blue histograms are for cE(w)s and cE(w/o)s, respectively. If most cEs are formed through tidal stripping from the nearest massive galaxy, we can expect a skewed distribution in which the majority of cEs are observed to lie in close proximity to the massive galaxy. However, our cE sample shows a wide range of projected distances from the luminous galaxy. While cEs show a peak close to the host galaxy (i.e., cE(w)s), it is interesting to note that a number of cEs also exist outside one $R_{\rm vir}$ of the luminous galaxy (i.e., cE(w/o)s). Approximately 22$\%$ of cEs are located beyond ten $R_{\rm vir}$ of the luminous galaxy. 

To obtain insights regarding the local environments of the cE(w/o)s, we also compared the distribution of cE(w/o)s with possible blue, star-forming dwarf galaxies (see the black open histogram in Fig.~\ref{fig:Envir}). Using the SDSS DR12 and GALEX GR7 UV data  \citep{Martin2005}, 6316 blue dwarf galaxies in the redshift range of $z < 0.05$ were chosen; these have the same luminosity range ($M_{g} > -18.7$ mag) as that of the cEs, but larger sizes ($R_{eff} >$ 600 pc) and bluer colors ($g-r < 0.3$ and $NUV-r < 2$). The nearest luminous galaxies of blue dwarf galaxies were defined with the same criteria as those used for the cEs. As shown in Fig.~\ref{fig:Envir}, most blue dwarf galaxies exist beyond one $R_{\rm vir}$ of the luminous galaxy. This is consistent with the findings that the gas-rich, star-forming dwarf galaxies preferentially lie at larger distances from the massive galaxy (e.g., \citealt{Mateo1998,Tolstoy2009,Weisz2011,Weisz2015} for the Local Group). \citet{Geha2012} also demonstrated that most of the dwarf galaxies located beyond four $R_{\rm vir}$ of a neighboring massive galaxy exhibit star formation (see arrow in Fig.~\ref{fig:Envir}). It is interesting to note that a large fraction of our cE(w/o) sample are overlapped with the distribution of blue dwarf galaxies in which approximately 63$\%$ of cE(w/o)s are located beyond four $R_{\rm vir}$. This could imply that most cE(w/o)s share low-density local environments with star-forming dwarf galaxies.

Additionally, we estimated the global environment of cEs using the group catalog of \citet{Tempel2014}. We defined the host structure surrounding a cE if the cE is located within two $R_{\rm vir}$ of the nearest group and the radial velocity difference between the cE and group is smaller than the velocity dispersion of the group. We used the number of galaxies in the structure of the group catalog to classify the global environments of cEs into a cluster, group, and field following \citet{Norris2014}: a cluster has more than 40 galaxies, a group has 10 -- 40 galaxies, and a field has fewer than 10 galaxies. Furthermore, regardless of the number of the structure, if a cE is located beyond two $R_{\rm vir}$ of the nearest structure, the environment of the cE was considered to be the field. In our cE sample, 23, 24, and 91 cEs are found in the cluster, group, and field, respectively.

Figure~\ref{fig:MS} shows the location of our cE sample (circles) in the size-stellar mass space relative to other galaxies, including early-type galaxies (E/S0s, green boxes), dEs (crosses), and cEs (stars) compiled from literature \citep{Bender1993,Ferrarese2006,Misgeld2008,Misgeld2009,Norris2014,Guerou2015}. The stellar masses of our cEs are measured using the relation between the SDSS g-r color and the stellar mass-to-light ratio based on the i-band luminosity \citep{Bell2003} assuming the initial mass function of \citet{Kroupa1993}. The dotted line represents the linear best fit to E/S0s. The dEs differ from E/S0s in that dEs show small size variations over the large mass range, whereas E/S0s exhibit a steeper distribution. It is clear that most of the cEs in our sample are found in the same region as that occupied by previously known cEs. cEs locate below the distribution of dEs at a given mass, characterized by systematically smaller sizes than the diffuse low-mass galaxies with comparable masses. They instead fall on the extension of the relation defined by more massive E/S0s or lie in between the relations of E/S0s and dEs \citep{Kormendy2009,Misgeld2011}. In Fig.~\ref{fig:MS}, we cannot see distinct different distributions between the cE(w)s (red circles) and cE(w/o)s (blue circles), implying that the size and stellar mass of cEs are in a similar range regardless of the environment.
 
\begin{figure*}
\vspace{3mm}
\includegraphics[width=\textwidth]{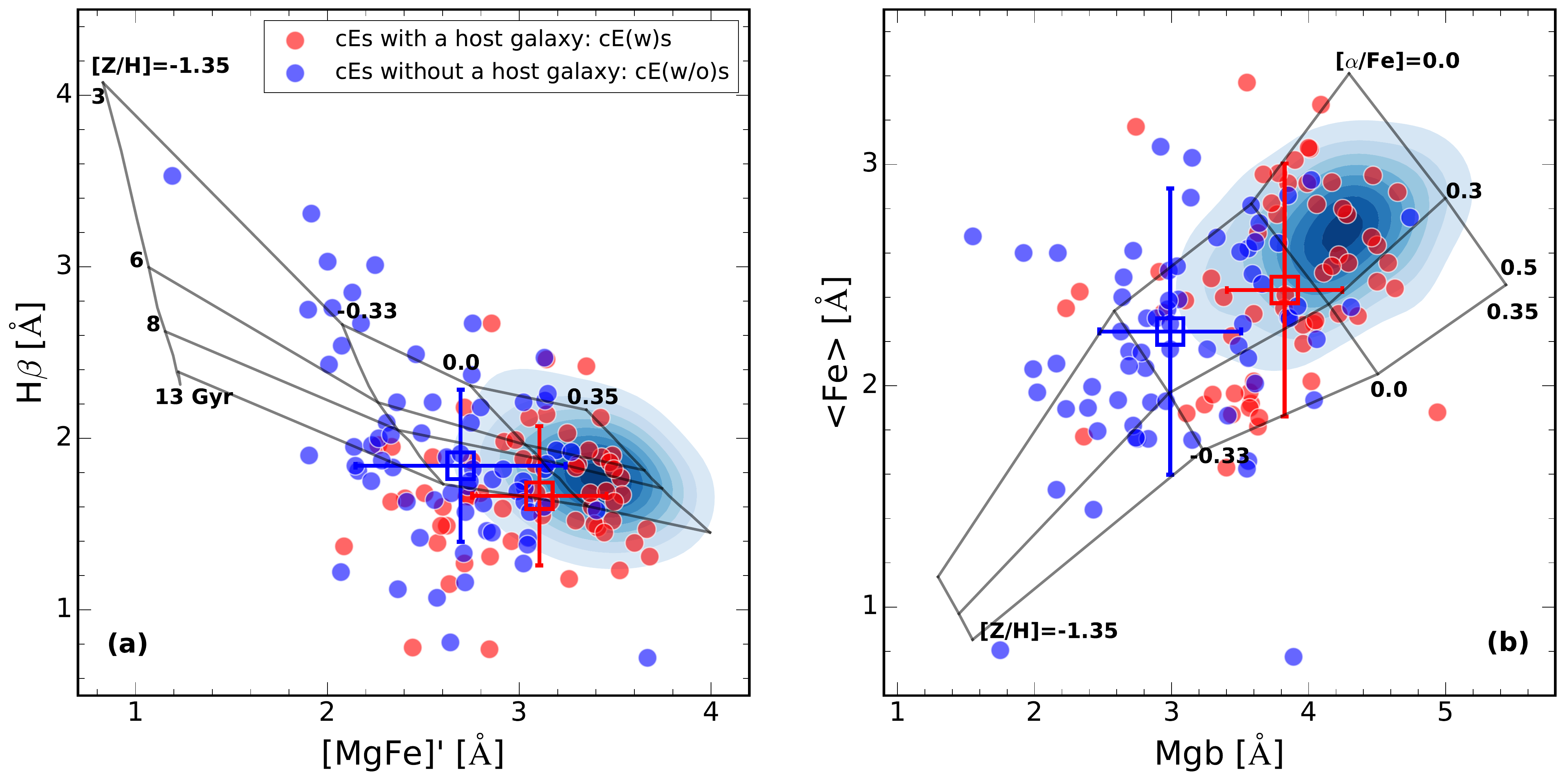}
\caption{\hb\ vs. \mgfe\ (a) and \fe\ vs. \mgb\ (b) diagrams of cEs. Red and blue circles indicate cE(w)s and cE(w/o)s, respectively. The median values of line indices of cE(w)s and cE(w/o)s with $1 \sigma$ errors are shown as red and blue squares, respectively. Contours denote the distribution of early-type galaxies extracted from the OSSY catalog \citep{Oh2011}. The model grids (solid lines) represent the SSP models for various values of age, [Z/H], and \afe\ given by \citet{Thomas2003} with \afe\ = 0.3 (a) and age = 10 Gyr (b).
}
\label{fig:model}
\end{figure*}

\subsection{Stellar Populations of cEs}

We investigated the stellar population properties of cEs based on \hb, \mgb, \fei, and \feii\ Lick indices extracted from the OSSY catalog \citep{Oh2011}. The OSSY catalog provides an improved database of absorption and emission line measurements for all SDSS galaxies at $z < 0.2$ using the \texttt{gandalf} line-fitting code \citep{Sarzi2006}.

Figure~\ref{fig:model} shows the observed line indices of cEs in comparison with the simple stellar population (SSP) model grid with various values of age, [Z/H], and \afe\ provided by \citet{Thomas2003}. The red and blue circles represent cE(w)s and cE(w/o)s, respectively. For comparison, we also present the distribution of early-type galaxies from the OSSY catalog \citep{Oh2011} as contours. In Fig.~\ref{fig:model}a, we present the distribution of cEs in the plane of \hb\ vs. \mgfe\footnote{\mgfe\,=\,$\sqrt{\mgb\,(0.72\,\times\,\fei\,+\,0.28\,\times\,\feii)}$ \citep{Thomas2003}} overlaid with the model grids with a fixed \afe\ value of 0.3. \hb\ is an age indicator and the composite index \mgfe\ is a good metallicity tracer \citep{Thomas2003}. While the majority of cEs show a similar [Z/H] distribution for early-type galaxies, a fraction of the cEs show smaller [Z/H] values. Most cEs also exhibit old ages similar to early-type galaxies. Interestingly, the metallicity distributions between the cE(w)s and cE(w/o)s appear different in the sense that cE(w)s are more biased toward higher \mgfe\ values than cE(w/o)s (see red and blue open squares for the median values of line indices of cE(w)s and cE(w/o)s, respectively).

Fig.~\ref{fig:model}b displays line indices of \fe\footnote{\fe\,=\,(\fei\,+\,\feii)/2 \citep{Gorgas1990}} and \mgb\ with the model grids for a representative age of 10 Gyr. It is clear that cEs show higher dispersion in [Z/H] values compared to early-type galaxies. Furthermore, the difference between the cE(w)s and cE(w/o)s in terms of their metallicities is prominent; cE(w)s are more metal-rich than cE(w/o)s (see red and blue open squares for the median values of line indices of cE(w)s and cE(w/o)s, respectively).

We derived the age, [Z/H], and \afe\ values of cEs by comparing the observed Lick indices with the SSP model grids of \citet{Thomas2003}. In the model grid of \hb$-$\mgfe, we need an \afe\ value for accurate estimation of the age and [Z/H] values because the dependence of this grid on \afe\ is not negligible. The \fe$-$\mgb\ grid also requires an age value for accurate estimation of [Z/H] and \afe\ values. Thus, to determine accurate age, [Z/H], and \afe\ values, we applied the iteration between two grids following the technique described in \citet{Puzia05}. For cEs located outside the model grids, we used the extreme grid value.

\begin{figure}
\vspace{3mm}
\centering
\includegraphics[width=3.3in]{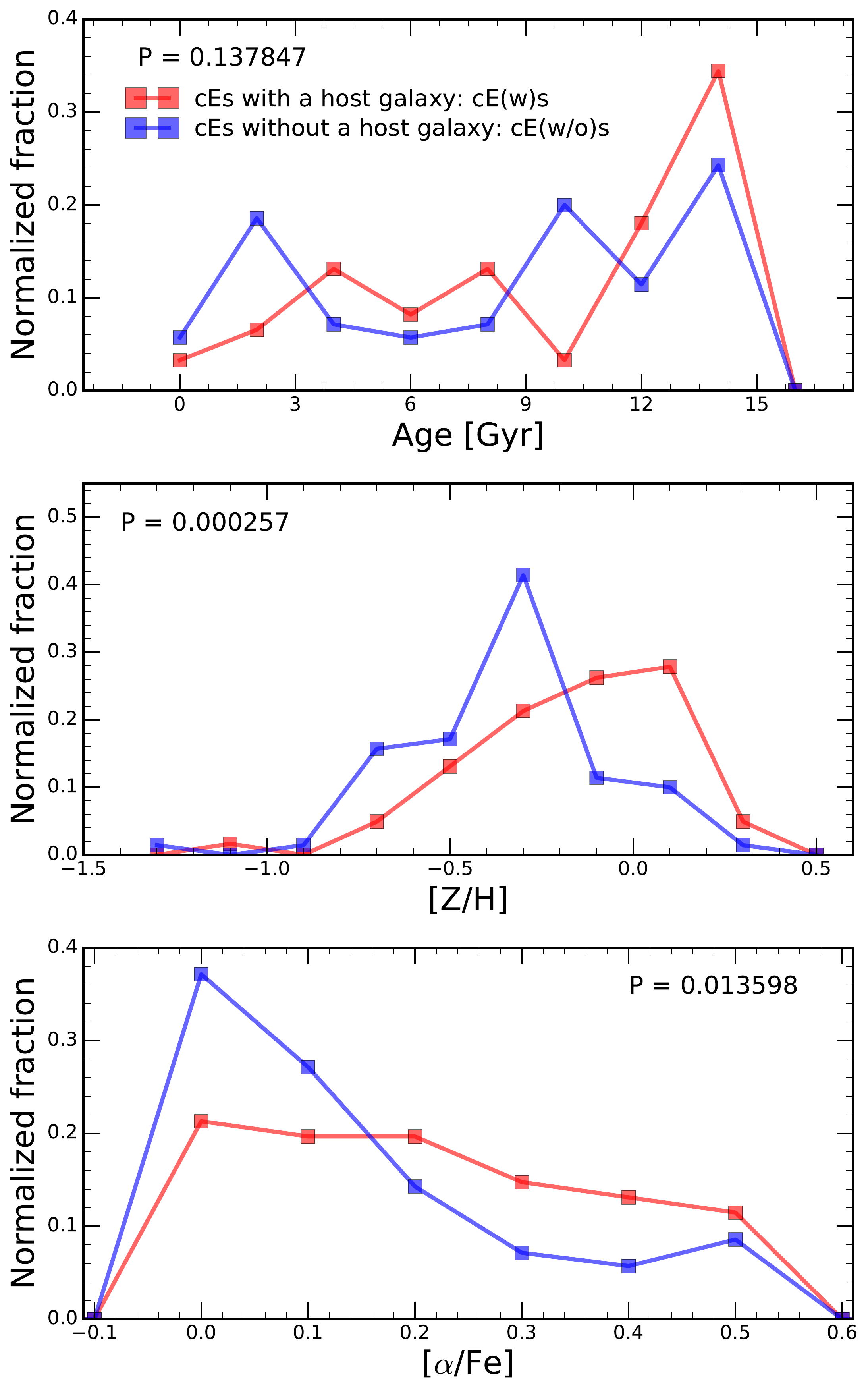}
\caption{Age (top), [Z/H] (middle), and \afe\ (bottom) distributions of cEs. The red and blue curves represent cE(w)s and cE(w/o)s, respectively. In each panel, the probability obtained from the K-S test is also given.}
\label{fig:histo}
\end{figure}

Figure~\ref{fig:histo} shows the distributions of derived age, [Z/H], and \afe\ values of cE(w)s (red curve) and cE(w/o)s (blue curve). The most notable feature is that the [Z/H] distribution of cE(w)s appears to be different from that of cE(w/o)s; cE(w)s show a skewed distribution toward higher [Z/H] values, whereas cE(w/o)s have a lower peak in the distribution. In the \afe\ distribution, cE(w/o)s exhibit a high fraction of relatively small \afe\ values, whereas cE(w)s show a rather flat distribution. As for the age distribution, a distinct difference is not shown between the cE(w)s and cE(w/o)s. We performed a K-S test to quantify the statistical significance of the differences in stellar population properties between the cE(w)s and cE(w/o)s. The test yields probabilities of $<$0.05, rejecting the null hypothesis of the same parent distribution between the cE(w)s and cE(w/o)s for [Z/H] and \afe. However, in the case of age, the probability is higher than 0.1, accepting the null hypothesis of the same underlying distribution. This indicates no statistically significant difference in the age distributions between the cE(w)s and cE(w/o)s.

\begin{figure}{}
    \vspace{3mm}
    \includegraphics[width=3.3in]{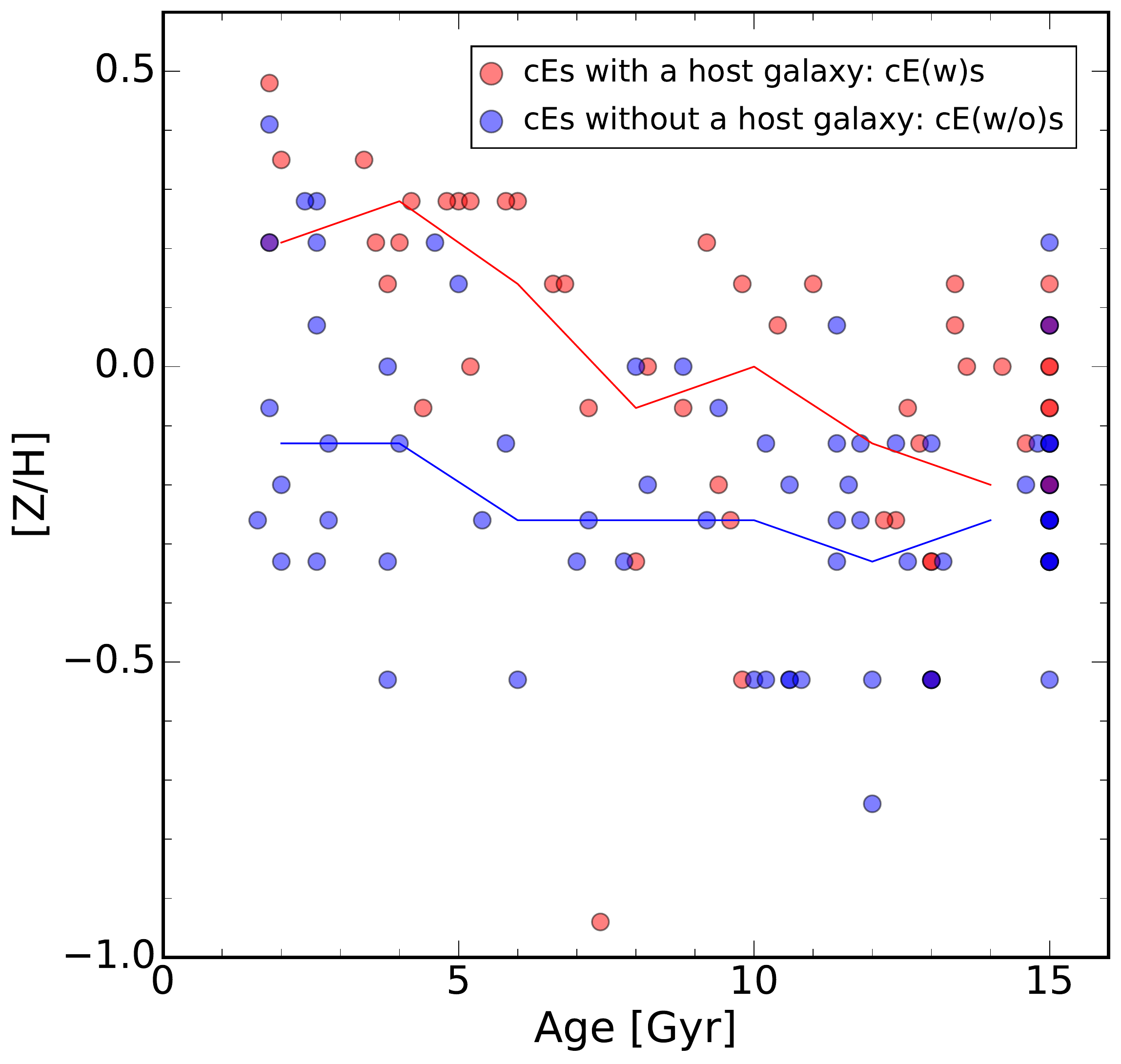}
    \centering
    \caption{[Z/H] vs. age distribution of cE(w)s (red circles) and cE(w/o)s (blue circles). The red and blue curves represent the mean loci of cE(w)s and cE(w/o)s, respectively.}
    \label{fig:AgeZH}
\end{figure}

In the metallicity vs. age distribution obtained by \citet{chilingarian2015}, there appears to be a difference in metallicities between cEs in isolated and dense environments \citep[see Fig. 2 of ][]{chilingarian2015}. Eight of the 11 isolated cEs appear to be slightly more metal-poor than the majority of their counterparts in dense environments at a given age. However, they concluded that the metallicities of their isolated cEs do not show a statistically significant difference from those of nonisolated cEs. In Figure ~\ref{fig:AgeZH}, we present the [Z/H] vs. age distribution of our cE sample. We also overplot mean loci of cE(w)s (red curve) and cE(w/o)s (blue curve) by calculating the running average values of [Z/H] along the age, where the bin size and the step size of the age are 4 Gyr and 2 Gyr, respectively. It is clear that the distribution of cE(w)s (red circles) is well separated from that of cE(w/o)s (blue circles) in that cE(w)s are systematically metal-rich than cE(w/o)s at any age. This result is in contrast to that obtained by \citet{chilingarian2015}, which could be attributed to the small sample size of their isolated cEs compared to ours.

\subsection{Mass-Metallicity Relation}

It is well known that metallicity correlates with stellar mass in the sense that high-mass galaxies show higher metallicity. In this context, it is interesting to investigate the mass-metallicity distribution of cEs regarding their possible connection with galaxies in the higher mass regime. Moreover, this would provide information on the mass of the progenitors, which can be used to constrain their formation. 

\begin{figure*}
\vspace{3mm}
\includegraphics[width=\textwidth]{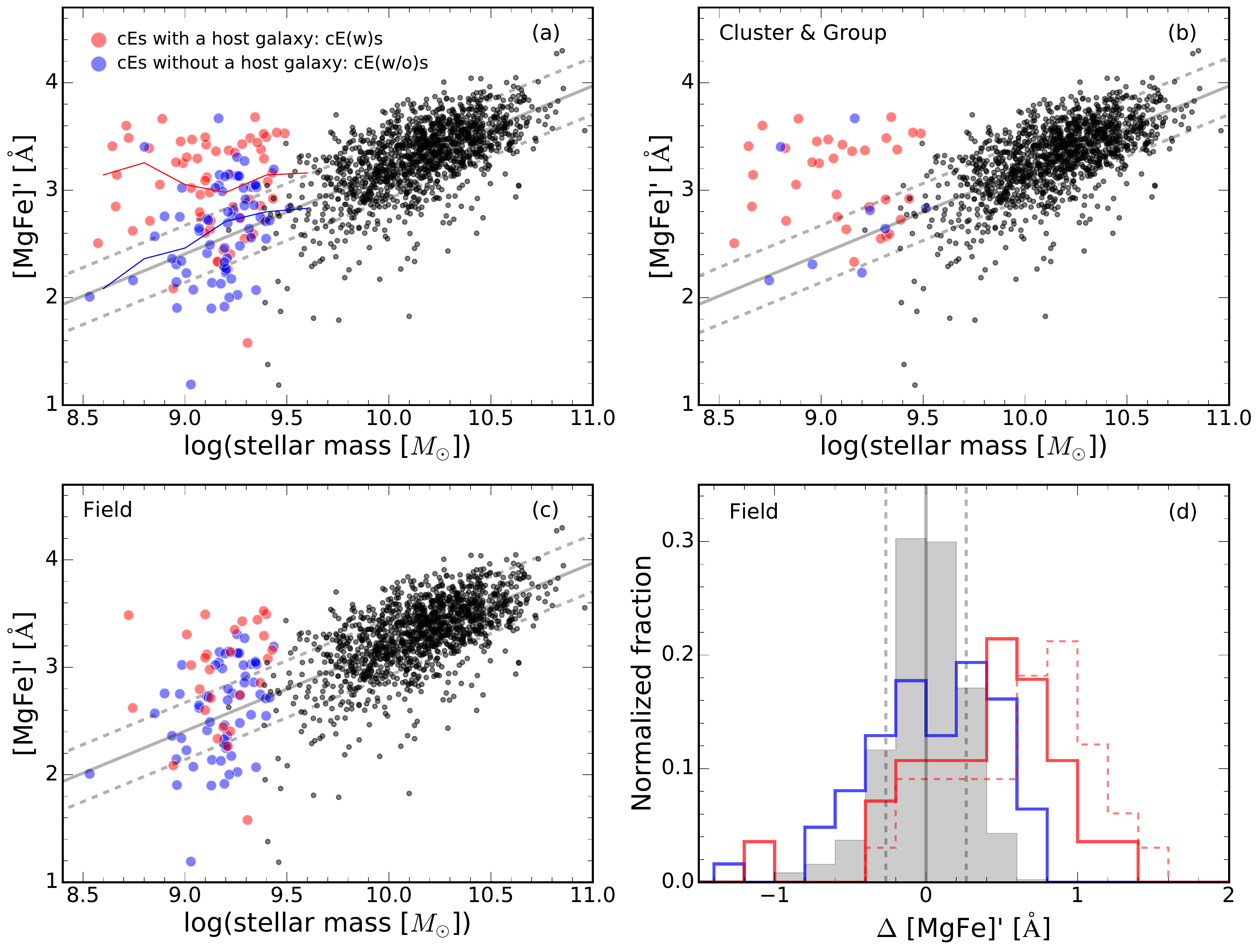}
\caption{(a) \mgfe\ vs. stellar mass distributions of cE(w)s (red circles) and cE(w/o)s (blue circles). Black small circles are early-type galaxies from the OSSY catalog \citep{Oh2011}. The solid line is the linear best fit for early-type galaxies and dotted lines denote its 1 $\sigma$ deviation. The red and blue curves represent the mean loci of cE(w)s and cE(w/o)s, respectively. (b) Same as (a), but for cEs in cluster and group environments. (c) Same as (a), but for cEs in the field environment. (d) Distributions of $\Delta$\mgfe, which is the difference between observed \mgfe\ values and expected ones from the mass-metallicity relation of early-type galaxies at a given mass, of cE(w)s (red histogram) and cE(w/o)s (blue histogram) in the field environment. The gray histogram is for early-type galaxies. The red dashed histogram denotes the distribution of cE(w)s in cluster and group environments. The vertical solid and dotted lines correspond to the linear best fit for early-type galaxies and its 1 $\sigma$ deviation in \mgfe\ vs. stellar mass distribution, respectively.
}
\label{fig:MassMgb}
\end{figure*}

In Figure~\ref{fig:MassMgb}a, we present the distributions of cE(w)s (red circles) and cE(w/o)s (blue circles) in the plane of \mgfe\ vs. stellar mass. For comparison, we also plot early-type galaxies from the OSSY catalog \citep[black small circles; ][]{Oh2011}. The stellar masses of the early-type galaxies are also measured using the relation between the SDSS g-r color and the stellar mass-to-light ratio based on the i-band luminosity \citep{Bell2003}. The solid line is the linear best fit for early-type galaxies and dotted lines denote its 1$\sigma$ deviation. We also overplot mean loci of cE(w)s (red curve) and cE(w/o)s (blue curve) by calculating the running average values of \mgfe\ along the stellar mass, where the bin size and the step size of the log(stellar mass) are 0.4 $M_{\odot}$ and 0.2 $M_{\odot}$, respectively. cEs exhibit a large range of \mgfe\ values at any given stellar mass compared to early-type galaxies. The most prominent feature is that cE(w)s have systematically higher \mgfe\ values than the majority of cE(w/o)s at a fixed mass. Furthermore, cE(w)s are located mostly above the +1$\sigma$ of the linear best fit line. In contrast, most cE(w/o)s well follow the relation corresponding to early-type galaxies and are well confined within 1$\sigma$ lines (also see the blue curve for the mean locus of cE(w/o)s). This implies that cEs in different local environments are clearly separated with respect to their mass-metallicity relation in that cEs associated with a host galaxy are more metal-rich than their isolated counterparts at a given mass.

Further, to examine the features of cEs with respect to the large-scale, global environment (see Sec. 3.1), we also present the distributions of cEs in dense (i.e., cluster and group, Fig.~\ref{fig:MassMgb}b) and sparse (i.e., field, Fig.~\ref{fig:MassMgb}c) environments. As expected, most ($80\%$) cEs residing in the cluster and group environments are associated with a host galaxy (i.e., cE(w)s). These galaxies show large deviations from the mass-metallicity relation of early-type galaxies. Most of the cEs in the field environment are those with no host galaxy (i.e., cE(w/o)s); these galaxies conform to the mass-metallicity relation of early-type galaxies. However, it is interesting to note that, contrary to our expectation, a non-negligible fraction ($31\%$) of cEs in the field environment also have a plausible host galaxy nearby. Furthermore, even in the field environment, the \mgfe\ vs. mass distribution of cE(w)s is also clearly different from that of cE(w/o)s, similar to that shown in dense environments.

In Fig.~\ref{fig:MassMgb}d, we present the distribution of $\Delta$\mgfe, which is the difference between the observed \mgfe\ values and expected ones from the mass-metallicity relation of early-type galaxies at a given mass of cE(w)s (red histogram) and cE(w/o)s (blue histogram) in the field environment. It clear that the $\Delta$\mgfe\ distributions between the cE(w)s and cE(w/o)s are different. While the $\Delta$\mgfe\ distribution of cE(w/o)s is very similar to that of early-type galaxies (gray histogram), the distribution of cE(w)s shows offsets from those of cE(w/o)s and early-type galaxies. This is confirmed by the statistical significance of the K-S tests; the probabilities of $\Delta$\mgfe\ distributions of the early-type galaxies vs. cE(w/o)s and of the early-type galaxies vs. cE(w)s are 0.051 and 0.001, respectively. Moreover, the distribution of cE(w)s in the field is very similar to that of cE(w)s in the cluster and group environments (red dashed histogram).

\section{DISCUSSION and CONCLUSIONS}
In this study, we explored the environmental dependence of the population properties of a large sample of cEs at z $<$ 0.05 using the SDSS DR12. Examination of the distance from a neighboring luminous galaxy suggests the existence of a distinct population of cEs that is beyond the $R_{\rm vir}$ of a luminous galaxy (i.e., cEs with no host galaxy; cE(w/o)s) in addition to the population that is located close to a luminous galaxy (i.e., cEs with a host galaxy; cE(w)s) (see Fig.~\ref{fig:Envir}). Our classification of cEs in this small-scale, local environment shows reasonable agreement with that for the large-scale, global environment. Most cEs in dense environments such as cluster and group are present in the vicinity of a host galaxy (see Fig.~\ref{fig:MassMgb}b). However, contrary to our expectation, some fraction of cEs in the field environment also show a plausible host galaxy nearby (see Fig.~\ref{fig:MassMgb}c and Fig.~\ref{fig:MassMgb}d). The existence of a non-negligible fraction of cE(w)s in the field environment suggests that the local environment associated with a nearby massive galaxy may also be essential for investigating the properties and formation of cEs depending on the environment.

Our most important result is that the metallicity distributions between the two cE subsamples, cE(w)s and cE(w/o)s, are significantly different. cE(w)s show a skewed distribution toward higher metallicities compared to that of cE(w/o)s (see Fig.~\ref{fig:histo} and Fig.~\ref{fig:AgeZH}), which is statistically confirmed by the K-S test (see Sec. 3.2). Most of the cE(w)s deviate from the mass-metallicity relation of early-type galaxies, being more metal-rich than metallicities that could be expected from their low masses (see Fig.~\ref{fig:MassMgb}a). Therefore, cE(w)s lie to the left (i.e., lower stellar mass) of early-type galaxies in the plane of the \mgfe\ vs. mass. This indicates that the metallicities of cE(w)s are rather similar to those of more massive early-type galaxies. However, the bulk of cE(w/o)s well follow the mass-metallicity relation of early-type galaxies in the low mass regime. Metallicity can be a measure of the original mass of the progenitor of a cE in terms of the mass-metallicity relation; this reinforces our conclusion that cE(w)s and cE(w/o)s are distinct classes of cEs.

The high metallicities of cE(w)s suggest that cE(w)s are initially larger, more massive galaxies but stripped down by a nearby host galaxy through a stripping process. Therefore, in the mass-metallicity distribution, high-mass progenitors of cE(w)s can move to remnant galaxies with lower stellar mass but preserving their high metallicities. According to the [Z/H] vs. mass relation determined by \citet{DeMasi2018} and mean mass of our cE(w)s $\sim 10^{9.25}M_{\odot}$, the progenitors of cE(w)s would initially have a mean stellar mass of 10$^{10.34}M_{\odot}$. This indicates that cE(w)s lost more than 90$\%$ of their original masses in the stripping process \citep[see also ][]{Chilingarian2009,Norris2014,Ferre2018}. On the other hand, the existence of cE(w/o)s and their stellar populations with lower metallicities underline the necessity of an additional channel for the formation of present-day cEs. In this case, cE(w/o)s might be rather bonafide low-mass classical early-type galaxies.


We also observed a possible difference in the \afe\ distributions between the cE(w)s and cE(w/o)s, although its significance is lower than that of metallicity (see Fig.~\ref{fig:histo}). The \afe\ distribution of cE(w/o)s is skewed toward lower values and cE(w)s have a slightly higher ($\sim$0.15 dex) mean value of \afe\ than cE(w/o)s. In general, \afe\ traces the star formation history (SFH), reflecting the efficiency and time-scale of star formation in galaxies \citep{Thomas2005}. Moreover, \afe\ correlates with the galaxy mass; rapid ($<$1 Gyr) and efficient star formation occurred in the massive galaxies, while low-mass galaxies experience an extended ($>$1 Gyr) and less efficient star formation (e.g., \citealt{Spolaor2010} and references therein). In this context, our result is naturally compatible with the mass-dependent SFH at the early star-forming epoch. A higher mean \afe\ of cE(w)s implies an SFH with rapid star formation in their massive progenitors. In contrast, a lower \afe\ of cE(w/o)s is indicative of an extended SFH with relatively long star formation time-scale (a few Gyr), which is consistent with genuine low-mass galaxies\citep[see also ][]{FerreMateu2013}. Therefore, the difference in \afe\ between the cE(w)s and cE(w/o)s is another evidence supporting the existence of different original masses of progenitors corresponding to their different formation channels.

 Our results strengthen the suggestion that cEs are mixtures of galaxies with two types of origins depending on their environment (e.g., \citealt{Ferre2018} and references therein). One contains remnants of larger galaxies with bulges tidally stripped by a neighboring massive host galaxy (i.e., stripped cEs). The other comprises real low-mass classical early-type galaxies in an isolated environment with no massive galaxy nearby (i.e., intrinsic cEs). On the observational side, these two scenarios cannot be reliably distinguished in parameter spaces (e.g., size vs. mass diagrams) owing to the similarity in the structural parameters of all cEs. However, we can strongly suggest that the metallicity of cEs may be one of the key parameters responsible for discriminating between rival formation scenarios of cEs, because there should be residual metallicity signatures pertaining to different origins. In combination with other diagnostic tools based on high-quality spectroscopic observations, we can further infer the evolutionary stages and progenitor types for a large sample of cEs in future works.

\acknowledgments
We are grateful to the anonymous referee for helpful comments and suggestions that improved the clarity and quality of this paper.
S.K. acknowledges support from Basic Science Research Program through the National Research Foundation of Korea (NRF) funded by the Ministry of Education (NRF-2019R1I1A1A01061237).
S.C.R. acknowledges support from the Basic Science Research Program through the National Research Foundation of Korea (NRF) funded by the Ministry of Education (2018R1A2B2006445). Support for this work was also provided by the NRF to the Center for Galaxy Evolution Research (2017R1A5A1070354). 
H.J. acknowledges support from the Basic Science Research Program through the National Research Foundation (NRF) of Korea (NRF-2019R1F1A1041086).
S.J.J. acknowledges support from Basic Science Research Program through the National Research Foundation of Korea (NRF) funded by the Ministry of Education (NRF-2020R1I1A1A01055595).
We would like to thank Editage (www.editage.co.kr) for English language editing.

\bibliographystyle{yahapj}
\bibliography{references}

\end{document}